# Zero-order ultrasensitivity: A study of criticality and fluctuations under the total quasi-steady state approximation in the linear noise regime

P.K. Jithinraj, Ushasi Roy, Manoj Gopalakrishnan*

*Department of Physics, Indian Institute of Technology (Madras), Chennai 600036, India*

## HIGHLIGHTS

- A push-pull enzyme-substrate system is ultrasensitive under enzyme saturation.
- Deterministic chemical rate equations are inadequate for small substrate populations.
- We adopt a probabilistic approach, starting from master equation.
- Fluctuations are estimated within the linear noise approximation.
- Analytical results are supported by stochastic simulations.

**ABSTRACT** Zero-order ultrasensitivity (ZOU) is a long known and interesting phenomenon in enzyme networks. Here, a substrate is reversibly modified by two antagonistic enzymes (a 'push-pull' system) and the fraction in modified state undergoes a sharp switching from near-zero to near-unity at a critical value of the ratio of the enzyme concentrations, under saturation conditions. ZOU and its extensions have been studied for several decades now, ever since the seminal paper of Goldbeter and Koshland (1981); however, a complete probabilistic treatment, important for the study of fluctuations in finite populations, is still lacking. In this paper, we study ZOU using a modular approach, akin to the total quasi-steady state approximation (tQSSA). This approach leads to a set of Fokker-Planck (drift-diffusion) equations for the probability distributions of the intermediate enzyme-bound complexes, as well as the modified/unmodified fractions of substrate molecules. We obtain explicit expressions for various average fractions and their fluctuations in the linear noise approximation (LNA). The emergence of a 'critical point' for the switching transition is rigorously established. New analytical results are derived for the average and variance of the fractional substrate concentration in various chemical states in the near-critical regime. For the total fraction in the modified state, the variance is shown to be a maximum near the critical point and decays algebraically away from it, similar to a second-order phase transition. The new analytical results are compared with existing ones as well as detailed numerical simulations using a Gillespie algorithm.

*Corresponding author at: Department of Physics, Indian Institute of Technology (Madras) Chennai 600036, India. Tel.: +91-44-22574894; Fax: +91-44-2257485.
*E-mail addresses*: jithin@physics.iitm.ac.in (P K Jithinraj), ushasi@physics.iitm.ac.in (Ushasi Roy), manojgopal@iitm.ac.in (Manoj Gopalakrishnan)

# 1. Introduction

Goldbeter and Koshland (1981, hereafter referred to as GK) first showed that reversible covalent modification (e.g. phosphorylation or methylation) of a protein (substrate), catalyzed by two enzymes, contains within it a mechanism equivalent to a molecular switch. This switch-like behavior emerges in the limit where the substrate concentration far exceeds the enzyme concentrations as well as their individual Michaelis constants, as a consequence of which the enzymes work in the 'zero-order' regime. In this regime, the net modification and de-modification rates, predicted by standard Michaelis-Menten kinetic equations, become effectively independent of concentration (hence called 'zero-order', as opposed to the first order regime, where the rates depend linearly on concentrations). The chemical equilibrium condition (which translates to a quadratic equation for the modified fraction when intermediates are neglected, and a cubic equation when they are not) predicts that the fraction of substrate in modified state is either none or all, in the limit of large substrate concentrations. Specifically, the solution of this equation displays the switch-like behavior described above as a function of the ratio $\alpha \equiv v_r R_0 / v_b B_0$, where $R_0$ and $B_0$ are the enzyme concentrations and $v_r$ and $v_b$ their conversion rates. The 'critical point' of this transition was shown to be at $\alpha = 1$, (hereafter referred to as the GK point) independent of the ratio of the Michaelis constants of the enzymes. (Note: Throughout this paper, we shall use the words critical point and criticality in connection with ZOU although, despite many similarities, it is not a thermodynamic phase transition in the strict sense).

The GK switch was studied in more detail by some authors (e.g., Berg et al., 2000; Qian, 2003; Elf and Ehrenberg, 2003; Bluthgen et al., 2006; Ciliberto et al., 2007; Gomez-Uribe et al., 2007; Ge and Qian, 2008; Pederson and Bersani, 2010; Xu and Gunawardena, 2012) and also extended in scope by others (Ortega et al., 2002; Samoilev et al., 2005; van Albada and ten Wolde, 2007; Szomolay and Shahrezaei, 2012) in more recent times. Notably, Berg et al. (2000) and later, Elf and Ehrenberg (2003) studied the fluctuations in the ultrasensitive module within some approximations (see discussions later) while Qian (2003), and later Ge and Qian (2008), identified ZOU as a temporal cooperativity phenomenon, mathematically similar to the better known allosteric cooperativity. In connection with ZOU, Bluthgen et al. (2006), Ciliberto et al. (2007) and Pederson and Bersani (2010) showed that the total quasi-steady state approximation (tQSSA), introduced by Borghans et al. (1996) and studied further by Tzafriri and Edelman (2004) is superior to the Briggs-Haldane standard quasi-steady state approximation (sQSSA) when enzyme and substrate concentrations are comparable, or when the former actually exceeds the latter (whereas in sQSSA, the free substrate concentration is the slow variable, tQSSA replaces it with sum of the concentrations of the free substrate and the intermediate complex).

ZOU has been shown to be relevant in a number of systems (LaPorte and Koshland, 1983; Meinke et al., 1986; Cimino and Hervagault, 1987; Casati et al., 1999; Melen et al., 2005; Kim and Ferrell, 2007). Theoretical studies of the ZOU have, by and large, followed a chemical rate equation based approach, which is of a purely "mean-field" nature, and most authors have ignored fluctuations altogether. However, given the similarity of ZOU to a thermodynamic phase transition, it is natural to expect that biochemical fluctuations will be large near the transition point, an issue addressed in detail by Berg et al. (2000). In their model, a finite number $N$ of substrate molecules were considered, out of which, say, $n$ are in modified state at any given point of time, the probability for which was denoted $P_n$. The transition rates for the processes $n \leftrightarrow n \pm 1$ were assumed to be of the standard Michaelis-Menten type (derived under the sQSSA), and this, in our opinion, is a weakness of the model. The analytical calculations were restricted to the extreme case of an infinitely large substrate concentration. More recently, Elf and Ehrenberg (2003) obtained estimates for fluctuations in ZOU under the LNA; however, similar to Berg et al.(2000), macroscopic rates derived under sQSSA were used in their calculations. We shall attempt to show, in this paper, how these limitations can be overcome by using a different approach with a more controlled limiting procedure. Wherever relevant, we will also provide comparisons of our results with the older ones.

The principal objective of this work is the construction of a fully stochastic formulation of a two-state covalent modification system showing ZOU, with a complete treatment of fluctuations. For this purpose, we consider the system as consisting of two weakly connected modules, each populated by unmodified and modified substrate molecules respectively, in the spirit of tQSSA (Borghans et al., 1996). Discrete master equations are constructed to describe the dynamics in each, which are then converted to continuum Fokker-Planck equations by second-order truncation of the respective Kramers-Moyal expansions (Gardiner, 2004). A set of well-defined approximations, valid within the assumptions of tQSSA and the requirements of ZOU, then leads to an effective one-dimensional Fokker-Planck equation for the modified substrate fraction (the total population in the second module). Rigorous and elegant mathematical expressions for the averages and fluctuations in the steady state follow in a straightforward manner, which are shown to compare well with the results of detailed numerical simulations, done using a Gillespie algorithm (Gillespie, 1977) The present formalism can be potentially extended to more complex systems like a many-state reversible modification network (e.g., receptor methylation and demethylation in *E.coli*).

## 2. Model and methods

2.1. Fokker-Planck equations from the master equation

We consider a cell of volume $V$, which contains $N$ substrate molecules $A$ at total concentration $A_0$ and two enzymes (which we shall call R and B) with total concentrations $R_0$ and $B_0$ respectively. The enzyme R binds to $A$ with association rate $k_+$ and reversibly converts it to the intermediate state $\tilde{A}$; the backward transition $\tilde{A} \to A + R$ occurs at rate $k_-$, while $\tilde{A}$ is irreversibly converted to the product (modified form of $A$, which we denote $A^*$) at rate $v_r$. Similarly, $A^*$ reversibly binds to B with rate $k'_+$ forming the second intermediate complex $\tilde{A}^*$, which dissociates to $A^*$ and B at rate $k'_-$. The complex $\tilde{A}^*$ is converted back to the original, non-modified form $A$ at a rate $v_b$. The reaction scheme is illustrated in Fig. 1. We further define the following dissociation constants for the enzyme-substrate binding: $K_r = k_-/k_+$ ; $K_b = k'_-/k'_+$. This is the original model studied by Goldbeter and Koshland (1981).

In the limit where the rates $v_r$ and $v_b$ are small in comparison with the rates of enzyme binding and dissociation, the above system functions as a combination of two weakly coupled modules, the $A - \tilde{A}$ system (Module 1) with $M_1 \equiv N(1-\xi)$ substrate molecules and the $A^* - \tilde{A}^*$ system (Module 2) with $M_2 \equiv N\xi$ molecules, where $\xi$ is the total fraction of substrate molecules in Module 2. In the limit where the turnover rates $v_r$ and $v_b$ are sufficiently small (see more discussions later in Sec. 3.4), the internal dynamics of these modules occur on a timescale much smaller than the one involving changes in $\xi$ itself; hence we may assume the two modules to be always in their steady states for each $\xi$. This is the essence of the tQSSA. The regime of validity of this scheme for irreversible Michaelis-Menten kinetics is discussed by Borghans et al. (1996) and its extension to reversible Michaelis-Menten kinetics was carried out by Tsafriri and Edelman (2004).

In module 1, the probabililily, $P_{M_1}(m_1)$, for $m_1$ number of molecules to be in $\tilde{A}$ state, satisfies the master equation:

$$\frac{\partial P_{M_1}(m_1,t)}{\partial t} = \omega_+(m_1-1)P_{M_1}(m_1-1) + \omega_-(m_1+1)P_{M_1}(m_1+1) - [\omega_+(m_1) + \omega_-(m_1)]P_{M_1}(m_1), \quad (1)$$

with rates $\omega_+$ and $\omega_-$ defined as below:

$$\omega_+(m_1) = (M_1 - m_1)k_+ R_f(m_1) \quad ; \quad \omega_-(m_1) = m_1 k_- , \quad (2)$$

where $R_f(m_1) \cong R_0 - m_1/V$ is the concentration of free R. Let us now define the fraction $x \equiv m_1/M_1$. For large enough $M_1$, we may treat $x$ as a continuous variable, and define the continuous probability distribution $\Phi_x(x) \cong \delta_x^{-1} P_{M_1}(m_1)$ and rates $\omega_\pm^x(x) \equiv \omega_\pm(m_1)$, where $\delta_x = M_1^{-1}$ is the 'step size' for $x$, which may be regarded as a small parameter. In this limit, the master equation may be converted to a continuum Fokker-Planck equation via the Kramers-Moyal expansion, i.e., by expanding the quantities $\Phi_x(x \pm \delta_x)$ and $\omega_\pm^x(x \pm \delta_x)$ in a Taylor series in $\delta_x$, up to terms that are $O(\delta_x^2)$. The resulting equation has the form of an equation of continuity (see Appendix A for details)

$$\partial_t \Phi_x = -\partial_x J_x , \quad (3)$$

where

$$J_x = v_x(x)\Phi_x - \partial_x(D_x(x)\Phi_x) \quad (4a)$$

is the probability current density, with coefficients $v_x(x)$ and $D_x(x)$ given by

$$v_x(x) = \delta_x(\omega_+^x - \omega_-^x) \quad ; \quad D_x(x) = \frac{\delta_x^2}{2}\left[\omega_+^x + \omega_-^x\right] \tag{4b}$$

with

$$\omega_+^x = M_1(1-x)k_+ R_f(\xi, x) \quad \text{and} \quad \omega_-^x = M_1 x k_- . \tag{5}$$

In the above equation, $R_f(\xi, x) = R_0 - A_0(1-\xi)x$. In steady state, following (3), we require $\partial_x J_x = 0$. In addition, since $x \in [0,1]$, we require that $J_x = 0$ at the boundaries, i.e., $x = 0, 1$. These conditions are consistent only if we impose $J_x = 0$ identically for all $x$. Using this condition in (4), we arrive at the following steady state solution for the probability distribution (Gardiner, 2004):

$$\Phi_x(x) = \frac{C}{D_x(x)} \exp\left(\int_0^x \frac{v_x(x')}{D_x(x')} dx'\right), \tag{6}$$

where $C$ is a normalization constant. It is convenient to evaluate this integral by the following method. We assume the existence of a number $\bar{x}$, which satisfies the following conditions:

$$v_x(\bar{x}) = 0, \quad v_x'(\bar{x}) < 0 . \tag{7}$$

where $v_x'(x) \equiv \partial_x v_x(x)$. After expanding the numerator inside the integral in a Taylor series around $x = \bar{x}$ and keeping only the first two terms, it follows that the probability distribution $\Phi_x(x)$ is approximated by a Gaussian:

$$\Phi_x(x) \approx \Phi_x(\bar{x}) \exp\left[\frac{v_x'(\bar{x})}{2D_x(\bar{x})}(x - \bar{x})^2\right] . \tag{8}$$

so that $\bar{x}$ becomes the mean value of $x$ and is given by (7), while its variance, following (8), is given by

$$\sigma_x^2 \equiv \langle (x - \bar{x})^2 \rangle = -\frac{D_x(\bar{x})}{v_x'(\bar{x})} \tag{9}$$

Note, from (4b), that while the `velocity' $v_x$ is $O(M_1^0)$, the diffusion constant $D_x$ is $O(M_1^{-1})$ and hence small. Therefore, the fluctuation in (9) is $O(M_1^{-1})$ and vanishes as $M_1 \to \infty$. It follows that the further terms in the original Taylor expansion are even smaller and can be neglected in the limit of large $M_1$. This is the essence of the LNA.

The results (8) and (9) also appear as the leading terms for the corresponding quantities in the more systematic System Size Expansion, introduced by van Kampen (2008). In this formalism, $\bar{x}$ is called a fixed point, and the second part of (7) expresses its stability. The comparisons between Kramers-Moyal (whose truncation at second order leads to the general nonlinear Fokker Planck equation) and System Size expansions have been the subject of much recent discussion in the literature, and it has been shown rigorously that the agreement between the two is limited to the lowest order of the expansion in both cases (Grima et al., 2011). An alternative approach is the Poisson representation method due to Gardiner (2004); see, e.g., (Thomas et al. 2010) for an application in enzyme networks.

The formalism for module 2 is done in the same way as for module 1. We first define $y$ as the fraction of enzyme-bound proteins in module 2, the probability distribution of which can again be shown to be a Gaussian, as in (8), with the transformations $v_x \to v_y$ and $D_x \to D_y$. Here, $v_y(y)$ and $D_y(y)$ are defined by relations analogous to (5), along with the substitutions $\omega_\pm^x \to \omega_\pm^y$ and $\delta_x \to \delta_y$ where $\delta_y = M_2^{-1}$, and the rates $\omega_+^y$ and $\omega_-^y$ are given by

$$\omega_+^y = M_2(1-y)k'_+ B_f(\xi, y) \quad , \quad \omega_-^y = M_2 k'_- y \,. \tag{10}$$

where $B_f(\xi, y) = B_0 - A_0 \xi y$. The rest of the calculations for module 2 are similar to those for module 1, with appropriate replacement of variables as discussed above.

The complete, joint probability distribution for $\xi$, $x$ and $y$ together may be denoted by $\Gamma_N(\xi, x, y; t)$. In the large $N$ limit, it is straightforward to see that this distribution satisfies the more general equation of continuity:

$$\partial_t \Gamma_N(\xi, x, y; t) = -\partial_\xi J'_\xi - \partial_x J'_x - \partial_y J'_y \quad , \tag{11}$$

with current densities $J'_\zeta = v_\zeta \Gamma - \partial_\zeta (D_\zeta \Gamma)$ for $\zeta = x, y, \xi$. Analogous to (5), in general, we have $v_\zeta(\zeta) = \delta_\zeta(\omega_+^\zeta - \omega_-^\zeta)$ and $D_\zeta(\zeta) = (\delta_\zeta^2/2)[\omega_+^\zeta + \omega_-^\zeta]$ with $\delta_\xi = N^{-1}$. The forward and backward rates for $x$ and $y$ are given by (5) and (10) respectively, while the corresponding quantities for $\xi$ are given by

$$\omega_+^\xi = N(1-\xi)v_r x(\xi) \quad , \quad \omega_-^\xi = N\xi v_b y(\xi). \tag{12}$$

Starting from the (nonlinear) Fokker-Planck equation (11) for the variables $(x, y, \xi)$, it is possible now to obtain hierarchical equations for the various moments and cross-correlation functions of these variables. However, our principal interest here is to derive explicit expressions for the variance of $\xi$, for which it is convenient to use a simplified formalism using tQSSA under appropriate conditions.

Let us define $P(\xi,t) \equiv \int_0^1 dx \int_0^1 dy \Gamma_N(\xi,x,y;t)$ as the probability distribution of $\xi$. Under the tQSSA, owing to the wide separation in time scales between the slow variable $\xi$ and the fast variables $(x,y)$ (see discussions later in Sec 3.4), the function $\Gamma_N(\xi,x,y;t)$ may be expected to reduce to the product

$$\Gamma_N(\xi,x,y;t) \cong P(\xi,t)\Phi_x(x,t)\Phi_y(y,t) \ . \tag{13}$$

In order to find the equation satisfied by $P(\xi,t)$, let us now integrate (11) over $x$ and $y$ after substituting (13), which yields

$$\partial_t P(\xi,t) = -\partial_\xi \overline{J}'_\xi \quad \text{where} \quad \overline{J}'_\xi = \int dx dy J'_\xi = \overline{v}_\xi P - \partial_\xi(\overline{D}_\xi P) \ . \tag{14}$$

In the above, we also imposed the boundary conditions on $J'_x$ and $J'_y$ for fixed $\xi$ (see discussion following (5)). Next, using (12), we find

$$v_\xi(\xi) = \delta_\xi\left(\overline{\omega}^\xi_+ - \overline{\omega}^\xi_-\right) \quad \text{and} \quad D_\xi(\xi) = \frac{\delta_\xi^2}{2}\left[\overline{\omega}^\xi_+ + \overline{\omega}^\xi_-\right] \ , \tag{15}$$

where the effective forward and backward rates for $\xi$ are given by

$$\overline{\omega}^\xi_+ = N(1-\xi)v_r \overline{x}(\xi) \text{ and } \overline{\omega}^\xi_- = N\xi v_b \overline{y}(\xi) \ . \tag{16}$$

(14) has the same form as (3), with (16) replacing the rates in (5), therefore the method of solution follows the same arguments, with $\xi$ replacing $x$ and $N$ replacing $M_1$ everywhere. In particular, the average $\overline{\xi}$ satisfies the equation $\overline{v}_\xi = 0$, i.e.,

$$\overline{\xi}\overline{y}(\overline{\xi}) = (1-\overline{\xi})v\overline{x}(\overline{\xi}) \ , \tag{17}$$

where $v = v_r/v_b$. Analogous to (7), (17) gives the van Kampen fixed point of the total substrate fraction $\xi$ present in module 2.

The variance in the total modified substrate fraction, following (9), is given by

$$\sigma_\xi^2 \equiv \left\langle (\xi - \overline{\xi})^2 \right\rangle = -\frac{\overline{D}_\xi(\overline{\xi})}{\overline{v}'_\xi(\overline{\xi})} \tag{18}$$

in the LNA. It is obvious from (15) that $\sigma_\xi^2 = O(N^{-1})$ for large $N$. Note that, analogous to (7), we require that $\overline{v}'_\xi < 0$ (i.e., the fixed point in (17) should be stable) in order for (18) to be valid.

2.2. Stochastic numerical simulations

We verified all our mathematical results using Monte Carlo simulations of the system. A Gillespie algorithm was used to implement the reaction scheme in Fig. 1. To choose a set of specific numerical values for the various parameters for the simulations, we consider the methylation-demethylation of chemotaxis receptors in the bacterium *Escherichia coli*. This choice is by no means obvious or unique, and is primarily motivated by our future goals of extending the formalism in this study to understand the biochemical noise in signal transduction in *E. coli*. There is, indeed, a close connection between the two-state covalent modification system studied here, and the Barkai-Leibler (1995) model of adaptation in *E. coli*. The parameter values used in the simulations are summarized in Table 1.

We emphasize that our principal results do not depend in any way on the specific parameter values used, except for the general requirements of validity of the tQSSA. The results of the simulations are discussed in the following sections, along with the corresponding mathematical results.

**3. Results**

3.1. Determination of $\bar{x}$ and $\bar{y}$

Using (2) and (5) in (7), we see that the average $\bar{x}$ satisfies the following quadratic equation:

$$\bar{x}^2 A_0(1-\xi) - \bar{x}[K_r + R_0 + A_0(1-\xi)] - R_0 = 0 \tag{19}$$

with solution

$$\bar{x} = \frac{K_r + R_0 + A_0(1-\xi) \pm \sqrt{[K_r + R_0 + A_0(1-\xi)]^2 - 4A_0 R_0(1-\xi)}}{2A_0(1-\xi)} . \tag{20}$$

The minus sign gives the physically relevant solution. The formalism can be extended in exactly the same manner to module 2 to give the following expression for $\bar{y}(\xi)$:

$$\bar{y} = \frac{K_b + B_0 + A_0\xi - \sqrt{[K_b + B_0 + A_0\xi]^2 - 4B_0 R_0 \xi}}{2A_0\xi} \tag{21}$$

(20) and (21), combined with (17) give a mathematical expression, in the form of a chemical balance equation to determine $\bar{\xi}$. Here, the effective forward and backward rates are given by (16), which replace the standard Michaelis-Menten rates

$$\omega_+^{MM}(\xi) = \frac{N(1-\xi)v_r R_0}{K_r + A_0(1-\xi)} \quad \omega_-^{MM}(\xi) = \frac{N\xi v_b B_0}{K_b + A_0\xi} . \tag{22}$$

derived under sQSSA (Briggs and Haldane, 1925).

3.2. Ultrasensitivity and the critical point

We will now provide a rigorous proof that the GK point is a critical point for the ultrasensitive switching between $\xi = 0$ and $\xi = 1$. We do so by showing first that under conditions of enzyme saturation, the mathematical equation describing the modified substrate fraction become quadratic, as opposed to the cubic equation derived by Goldbeter and Koshland (1981) under more general, non-saturating conditions (reproduced in Appendix B). This simplifies the analysis considerably. The reader is also referred to Xu and Gunawardena (2012) for a powerful geometric approach which directly works with the cubic equation.

We start by conveniently expressing (19) and (20) in the following forms:

$$\bar{x}(\xi) = \frac{1}{2} + \frac{r + \kappa_r}{2(1-\xi)} - \frac{1}{2} f(r, \kappa_r, 1-\xi) \;;\; \bar{y}(\xi) = \frac{1}{2} + \frac{b + \kappa_b}{2\xi} - \frac{1}{2} f(b, \kappa_b, \xi) , \qquad (23)$$

where $r = R_0/A_0$, $b = B_0/A_0$, $\kappa_r = K_r/A_0$ and $\kappa_b = K_b/A_0$ are a set of dimensionless numbers which become vanishingly small as $A_0 \to \infty$, while the function $f(x, y, z) = \sqrt{1 + 2(y-x)/z + (x+y)^2/z^2}$. In the limit of large $A_0$, the variables $x$ and $y$ become small, and it is therefore useful to carry out a binomial expansion of the square root in the above expression, which yields the limiting form $f(x, y, z) \approx 1 + (y-x)/z$. It then follows that

$$\bar{x}(\xi) \approx \frac{r}{1-\xi}\left(1 - \frac{\kappa_r}{1-\xi}\right) + \frac{C_r}{4(1-\xi)^3} \quad \text{and} \quad \bar{y}(\xi) \approx \frac{b}{\xi}\left(1 - \frac{\kappa_b}{\xi}\right) + \frac{C_b}{4\xi^3} . \qquad (24)$$

where $C_r = \kappa_r^3 - r^3 + r\kappa_r(\kappa_r - r)$ and $C_b = \kappa_b^3 - b^3 + b\kappa_b(\kappa_b - b)$. Substitution of (24) (after neglecting the last terms in both, in the large $A_0$ limit) into (16) yields the following quadratic equation for $\bar{\xi}$.

$$(1-\alpha)\bar{\xi}^2 + [\alpha(1-\kappa_r) - (1+\kappa_b)]\bar{\xi} + \kappa_b = 0 \text{ where } \alpha = \nu r/b , \qquad (25a)$$

with solution

$$\bar{\xi}_{nc} = \frac{(1-\alpha) + (\kappa_b + \alpha\kappa_r) - \sqrt{[1-\alpha+\kappa_b+\alpha\kappa_r]^2 - 4(1-\alpha)\kappa_b}}{2(1-\alpha)} \qquad (25b)$$

where the subscript is an abbreviation for 'near-critical'. (25a) effectively replaces the cubic equation derived by Goldbeter and Koshland (1981) under the sQSSA, in the vicinity of the critical point, in the limit of large $A_0$.

The solution in (25b) is discontinuous across the point $\alpha = 1$, which we shall call the critical point. In terms of the original variables, this is equivalent to $R_0 = B_0/\nu \equiv R_c$, which is the same as the GK point (compare (25b) with (26b)). The limiting behavior of $\bar{\xi}$ in the limit of large $A_0$, in the sub-critical ($R_0 < R_c$) and super-critical ($R_0 > R_c$) regimes, obtained from an analysis of (25b), are given below:

$$\bar{\xi}_{nc} \approx \frac{R_c}{(R_c - R_0)} \frac{K_b}{A_0} \qquad R_0 < R_c, A_0 \to \infty \qquad (26a)$$

$$\bar{\xi}_{nc} \approx 1 - \frac{R_0}{(R_0 - R_c)} \frac{K_r}{A_0} \qquad R_0 > R_c, A_0 \to \infty \qquad (26b)$$

It is obvious that (23-25) work best in the vicinity of the critical point. This is because the binomial approximation leading to (24) from (23) is valid when $\bar{\xi} \gg \max(b, \kappa_b)$ and $1 - \bar{\xi} \gg \max(r, \kappa_r)$. Using the expressions (26) in the above conditions then leads to $|R_0 - R_c| \ll R_c$ as a self-consistency requirement for (26a) and (26b) to be valid.

(26a) and (26b) clearly show that the conditions $K_b \ll A_0$ and $K_r \ll A_0$ are required for the transition to be sharp, for, in these limits, $\bar{\xi} \to 0$ and $\bar{\xi} \to 1$ respectively in the sub-critical and super-critical regime, respectively. At the critical point, putting $\alpha = 1$ in (25a), we find

$$\bar{\xi}(R_c) \equiv \bar{\xi}_c = \frac{K_b}{K_r + K_b}, \qquad (27)$$

independent of $A_0$.

3.3 Comparison with the results of sQSSA

It is interesting that the conventional sQSSA (when the intermediates are neglected) also predicts a quadratic equation similar to (25a), which, in our notation, may be written as

$$(1-\alpha)\bar{\xi}^2 + [\alpha(1-\kappa_b) - (1+\kappa_r)]\bar{\xi} + \alpha\kappa_b = 0, \quad \text{(sQSSA)} \qquad (28a)$$

the solution of which is

$$\bar{\xi}_{GK} = \frac{(1-\alpha) + (\kappa_r + \alpha\kappa_b) - \sqrt{[1-\alpha+\kappa_r+\alpha\kappa_b]^2 - 4\alpha(1-\alpha)\kappa_b}}{2(1-\alpha)}. \qquad (28b)$$

Here, the subscript GK refers to Goldbeter and Koshland, who first derived (28a). The limiting behaviors of $\bar{\xi}_{GK}$ in the limit of large $A_0$ are as follows:

$$\bar{\xi}_{GK} \approx \frac{R_0}{(R_c - R_0)} \frac{K_b}{A_0} \qquad R_0 < R_c, A_0 \to \infty \qquad (29a)$$

$$\bar{\xi}_{GK} \approx 1 - \frac{R_c}{(R_0 - R_c)} \frac{K_r}{A_0} \qquad R_0 > R_c, A_0 \to \infty \qquad (29b)$$

Note that (29) may also be directly obtained from (26) through the exchange $R_0 \leftrightarrow R_c$ in the numerators. Comparison of (26) and (29) shows that the results of sQSSA (with intermediates neglected) and tQSSA agree close to the critical point ($R_0 \to R_c$), in the limit of large $A_0$. In the next section, we show that both these conditions are favorable for large separation of time scales of intra-modular and inter-modular dynamics, itself is a sufficient condition for validity of tQSSA.

3.4 Effective Michaelis-Menten rates from a rational approximation

We shall now discuss the connections between the effective modification/de-modification rates as emerging from the present approach and the conventional Michaelis-Menten rates derived under the sQSSA. The limiting behavior of $\bar{x}$ and $\bar{y}$ for small and large $R_0$ may be used to construct rational function approximations for them, which are easier to work with, compared to the exact quadratic forms in (19) and (20). We note that when $R_0 << K_r + A_0(1-\xi)$, the square root in (19) may be expanded binomially, and keeping the first two terms in the expansion yields

$$\bar{x} \approx \frac{R_0}{K_r + A_0(1-\xi)} \quad \text{when} \quad R_0 << K_r + A_0(1-\xi) \tag{30}$$

In the opposite limit (i.e., $R_0 >> K_r + A_0(1-\xi)$), we note that $\bar{x} \to 1$. The following rational function approximation captures these two asymptotic behaviors correctly:

$$\bar{x}_e(\xi) \approx \frac{R_0}{R_0 + K_r + A_0(1-\xi)} \tag{31a}$$

Similarly, a corresponding approximation formula may be constructed for $y(\xi)$, starting with (20), i.e.,

$$\bar{y}_e(\xi) \approx \frac{B_0}{B_0 + K_b + A_0 \xi} \ . \tag{31b}$$

Rational approximations of the form in (30) (sometimes referred to as Padè approximants ) have been used in earlier papers (e.g., Borghans et al., 1996; Ciliberto et al., 2007; Gomez-Uribe et al., 2007)We shall call them Effective Michaelis-Menten (EMM) rates in this paper, and denote the corresponding expressions with the symbol $e$ as subscript or superscript. These rates, when used in (22), result in the following effective forward and backward rates for the dynamics of $\xi$:

$$\omega_+^e(\xi) = \frac{N(1-\xi)v_r R_0}{R_0 + K_r + A_0(1-\xi)} \qquad \omega_-^e(\xi) = \frac{N\xi v_b B_0}{B_0 + K_b + A_0 \xi} \ , \tag{32}$$

which may be compared with (22). The expressions in (22) and (31) agree in the limits $R_0 \ll K_r$ and $B_0 \ll K_b$. The expressions in (30), when used in (16), yield the following approximate quadratic equation for $\bar{\xi}$ (Gomez-Uribe et al., 2007):

$$(\alpha-1)\xi^2 + [r+\kappa_r+\alpha(b+\kappa_b)+1-\alpha]\xi - \alpha(b+\kappa_b) = 0 \tag{33a}$$

the solution of which is given below:

$$\xi_e = \frac{-[r+\kappa_r+\alpha(b+\kappa_b)+1-\alpha] + \sqrt{[r+\kappa_r+\alpha(b+\kappa_b)+1-\alpha]^2 + 4\alpha(\alpha-1)(b+\kappa_b)}}{2(\alpha-1)} . \tag{33b}$$

In the limit of large $A_0$, this expression takes the following limiting forms:

$$\bar{\xi}_e \approx \frac{R_0(B_0+K_b)}{A_0(R_c-R_0)} \qquad \text{when} \quad R_0 < R_c \text{ and} \tag{34a}$$

$$\bar{\xi}_e \approx 1 - \frac{B_0(R_0+K_r)}{\nu A_0(R_0-R_c)} \qquad \text{when} \quad R_0 > R_c . \tag{34b}$$

The critical point $R_c$ has been defined in the last section. Note that while the solutions given in (33) are, in general, different from the near-critical expressions in (27), they become identical in the limits $R_0 \ll K_r$, $B_0 \ll K_b$ and $R_0 \to R_c$. The last condition conforms to our expectation that (27) is valid close to the critical point, while (33) may be expected to work away from it. Numerical results, discussed below, confirm these expectations.

A set of general conditions to be satisfied for the validity of tQSSA may now be derived by combining (16) with (31) and (34). From (16), the inter-modular transport of molecules is governed by the effective rates (per molecule) $v_r \bar{x}(\xi)$ and $v_b \bar{y}(\xi)$, therefore the applicability of tQSSA requires that both these rates are small in comparison with the enzyme binding and dissociation rates within the modules. Therefore, in general, we require that $\max(v_r \bar{x}, v_b \bar{y}) \ll \min(T_1^{-1}, T_2^{-1})$ where $T_1$ and $T_2$ are the time scales of equilibration in modules 1 and 2 respectively. For fixed $v_r$ and $v_b$, provided neither is too large in comparison with the (effective) forward and backward rates within the modules, this condition is satisfied if $\bar{x}, \bar{y} \ll 1$. (See, eg., Tsafriri and Edelman (2004) for a detailed discussion on the conditions for the magnitudes of $v_r$ and $v_b$ in the context of a *reversible* Michaelis-Menten reaction) This implies that a sufficient condition for the validity of tQSSA is that the fraction of molecules in the enzyme-bound intermediate state is small, which would mean that sQSSA applies equally well. Using (26), (29) or (34) in (31) shows that the condition is uniformly satisfied whenever $R_0/A_0 \ll 1$, $B_0/A_0 \ll 1$ and $R_0 \to R_c$. Therefore, we conclude that tQSSA applies well under conditions of enzyme saturation (which is a necessary requirement for ZOU) and when the system is close to the critical point. Furthermore, under these same conditions, its results should match that of sQSSA (with intermediate complexes neglected). The last point was already noted in our discussions following (29) above.

Coming now to numerical simulations, Fig. 2 shows the averages $\bar{x}$ and $\bar{y}$ plotted as functions of $R_0$. It is clearly seen that as the critical point is crossed, $\bar{x}$ shows a sudden rise from near-zero values to 1, while $\bar{y}$ drops from a large to small value. These observations lead us to interpret the ultrasensitive transition as a traffic jam: for small $R_0$, $\bar{x}$ is small, i.e., only a small number of substrate proteins are present in the first excited state, which then proceed (slowly) to the modified state and consequently processed to the second excited state by B-enzymes. As $R_0$ increases, a larger number of substrate molecules arrive in the modified state, but the B-enzymes (at fixed concentration) now do not have sufficient numbers to process them forward. As a result, there is a sharp drop in $\bar{y}$ with increase in $R_0$. Fig. 3b supports this conjecture, see the next paragraph. It is also interesting to note that both $\bar{x}$ and $\bar{y}$ are small near the critical point, in agreement with our earlier arguments. Two analytical results are also shown for comparison, corresponding to the EMM approximation in tQSSA (31) and the sQSSA prediction obtained from the GK cubic equation (B.6).

In Fig. 3a, we plot the average modified fraction as a function of $R_0$ for various concentrations $A_0$. The plots clearly show the sharpening of the transition as $A_0$ increases. A comparison of numerical data with the EMM solution (33b) is also shown in the inset of the figure. In Fig. 3b, the rates $v_r$ and $v_b$ were increased and reduced 10-fold, in comparison to their reference values in Fig. 3a. We observe that reduction of the rates has almost no discernible effect on the curve, but the transition is much weakened when the rates are increased. Intuitively, this is to be expected, as a high value of $v_b$ ensures that accumulation of substrate in module 2 is reduced for given $A_0$; however, as the arriving numbers increase even further with increasing $A_0$, accumulation also increases. Therefore, the net effect of an increase in the final conversion rates is to reduce the sharpness of the transition for given $A_0$.

Fig. 4a shows comparisons between the numerical data in Fig. 3b and theoretical expressions obtained by numerically solving (a) the van Kampen fixed point equation (17) and the GK cubic equation (B.4) using the *Mathematica 7.0* (Wolfram Research Inc, 2008) for two different sets of turnover rates with fixed ratio. When the turnover rates are large (main figure), the cubic equation, derived using sQSSA works well. This is to be expected, since the equilibration of the intermediate complexes with respect to the substrate, a requirement for sQSSA, is very fast when the turnover rate is large. The fixed point equation (17), being a function of the ratio of the turnover rates, is the same for the different curves in Fig. 3b and fails when the individual rates become too large. For smaller turnover rates (inset) the two theoretical curves are indistinguishable from each other and both agree splendidly with the simulation data.

Fig. 4b shows a similar comparison, but with data at a higher substrate concentration compared to Fig. 4a and with the predictions of various quadratic approximations, i.e., the near-critical expression (25b) the GK quadratic approximation (28b) and the EMM result (33b). As expected, the near-critical expressions work well near the critical point, but fail away from it, where EMM approach appears to work well.

## 3.5 Fluctuations in $\xi$ and power-law divergence near the critical point

We now focus on deriving expressions for the variance in $\xi$. In terms of $\bar{x}(\bar{\xi})$ and $\bar{y}(\bar{\xi})$, we have, for the variance, from (17),

$$\sigma_\xi^2 = \frac{1}{2N}\left\{\frac{\bar{\xi}\bar{y}(\bar{\xi}) + v(1-\bar{\xi})\bar{x}(\bar{\xi})}{\bar{\xi}\bar{y}'(\bar{\xi}) + \bar{y}(\bar{\xi}) - v\left[(1-\bar{\xi})\bar{x}'(\bar{\xi}) - \bar{x}(\bar{\xi})\right]}\right\} \quad (35)$$

In the large $A_0$ limit, and in the vicinity of the critical point, we may approximate $\bar{x} \approx r/(1-\bar{\xi})$ and $\bar{y} \approx b/\bar{\xi}$ from (24). After substitution of these expressions in (35), we find

$$\sigma_\xi^2 \cong \frac{1}{2N}\left[\frac{b + vr}{b\kappa_b\bar{\xi}^{-2} + vr\kappa_r(1-\bar{\xi})^{-2}}\right] \quad (36)$$

For $R_0 < R_c$, $\bar{\xi} \to 0$ as $A_0 \to \infty$, and the first term in the denominator dominates over the second, while for $R_0 > R_c$, the second term dominates, since $\bar{\xi} \to 1$ in this limit. In the next step, we substitute the near-critical expressions in (27) into (35), which leads to

$$\sigma_\xi^2 \cong \frac{(R_c + R_0)K_b R_c}{2VA_0^2(R_c - R_0)^2} \quad \text{when } R_0 < R_c, \quad (37a)$$

$$\sigma_\xi^2 \cong \frac{(R_c + R_0)K_r R_0}{2VA_0^2(R_0 - R_c)^2} \quad \text{when } R_0 > R_c \quad (37b)$$

and finally, using (28) in (35), we find

$$\sigma_\xi^2 = V^{-1}\frac{K_r K_b}{(K_r + K_b)^3} \quad \text{when } R_0 = R_c. \quad (37c)$$

In the above equations, we have made the substitution $N = A_0 V$, where $V$ is the cellular volume. In the special case $K_r = K_b = K$, we find $\sigma_\xi^2 = (8KV)^{-1}$ at the critical point.

Away from the critical point, it is more appropriate to use the rational expressions in (30) in (34), which gives

$$\sigma_\xi^2 \cong \frac{1}{2N}\left(\frac{b\bar{\xi}}{b\bar{\xi} + \kappa_b + \bar{\xi}} + \frac{vr(1-\bar{\xi})}{r + \kappa_r + 1 - \bar{\xi}}\right) \Big/ \left(\frac{b(b + \kappa_b)}{(b + \kappa_b + \bar{\xi})^2} + \frac{vr(r + \kappa_r)}{(r + \kappa_r + 1 - \bar{\xi})^2}\right) \quad (38)$$

(38) needs to be combined with (25b), (28b) or (33b) to express the variance as a function of $R_0$.

Fig. 5 shows the variance in $\xi$ plotted as a function of $R_0$ for different $A_0$, with comparison between simulations and the expression in (38). For best fit, we used $\bar{\xi}$ as measured in simulations (data in Fig. 3a) rather than the approximate analytical expressions. In general, we find good agreement between (38) and simulation data, except in the vicinity of the critical point and for small $A_0$. The observed disagreement in the latter case likely indicates the inadequacy of LNA at small substrate concentrations (see also discussions in the next section). Fig. 6 shows the variance at the critical point itself, as a function of $A_0$, compared with the theoretical prediction in (37c).

Fig. 7 shows the variance, at various concentrations $A_0$, plotted as a function of $A_0(R_c - R_0)$ for $R_0 < R_c$ (Fig. 7a) and as a function of $A_0(R_c - R_0)/R_0$ for $R_0 > R_c$ (Fig. 7b). The different forms of the scaling variables on the x-axis were suggested by the expressions (37a) and (37b) respectively. In general, we observe an algebraic relation between the variance and the appropriate scaling variable with a power -2. This behavior is reminiscent of the divergence of magnetic susceptibility $\chi$ (proportional to fluctuations in magnetization) near the critical point in a paramagnet-ferromagnet phase transition, i.e., $\chi \sim |T - T_c|^{-\gamma}$ with $\gamma$ being a critical exponent. More discussions on the analogy between ZOU and the liquid-gas phase transition in classical fluids may be found in Berg et al. (2000).

Finally, Fig. 8 shows that, away from criticality, the fluctuations decay proportional to $A_0^{-2}$ at fixed $R_0$ in both sub-critical and super-critical regimes, in agreement with (37a) and (37b).

## 4. Discussion and Conclusions

The Goldbeter-Koshland (GK) switch mechanism in reversible covalent modification continues to generate a great deal of interest among biologists, as its only requirement is the presence of an excess of substrate molecules over the modifying enzymes. Interestingly, mathematical calculations show that the sharpness of the response is determined only by this imbalance in substrate-enzyme concentrations, and, in principle, could be increased without limit. In effect, this means that, under appropriate conditions, such systems could achieve a level of sensitivity equivalent to allosteric enzymes with extremely high Hill coefficients. Goldbeter and Koshland (1981) named this amplification zero-order ultrasensitivity (ZOU), since its realization requires that the enzymes work in the `zero-order' regime, where they are saturated with respect to the substrate.

The work by Berg et al. (2000) discussed for the first time the intrinsic fluctuations in the GK ultrasensitive switch. These authors pointed out the similarities between ultrasensitivity and a thermodynamic phase transition (e.g. liquid-gas) and showed that the sensitivity of response, and hence fluctuations are maximized at the switching point. However, the analytical results were restricted to the limit $K/A_0 \to 0$, where $K$ represent the Michaelis constants. We show in Appendix C that this is a serious limitation of their model, as the leading terms in the expressions for the average as well as fluctuations of the modified substrate fraction are $O(K/A_0)$. We have also been able to reproduce the analytical results in Berg et al. (2000) under the same assumptions, but believe that our main results derived using the system size expansion technique are superior for the above reason.

We have not discussed the dynamics of the model in this paper, because we feel that this subject requires a careful and detailed treatment on its own (see also discussions in Berg et al. (2000) and Gomez-Uribe et al. (2003)). In particular, it is interesting to observe from (16) and (24) that, close to the critical point, the rates $\bar{\omega}_+^\xi$ and $\bar{\omega}_-^\xi$ become independent of $\xi$ in the large $A_0$ limit, and therefore, (14) that describes its dynamics reduces to a conventional diffusion equation with constant drift and diffusion terms. It would be interesting to develop a description of the dynamics of ZOU, with the random walk-like dynamics near the critical point as a starting point.

It is pertinent to make a few comments here about the limitations of our study. The tQSSA on which we based our calculations is useful in deriving explicit expressions for the fluctuations in the total substrate concentration in the modified state. We also identified the parameter regimes where tQSSA and the conventional sQSSA match in their results, and interestingly, these required conditions are found to favor ZOU. Although the present procedure was found to be sufficient to obtain reliable estimates of fluctuations, at least in the parameter regimes studied here, it may be worthwhile to carry out a multivariate systematic system size expansion in $x, y$ and $\xi$, which would yield more general results for the mean and variance for each of these variables, as well as their cross-correlation coefficients. In the leading order, the averages of $x, y$ and $\xi$ under this formalism coincide with the van Kampen fixed point in the $x-y-\xi$ space (e.g., (7) and (17)), and therefore match with tQSSA, but higher order corrections are likely to be different because of the appearance of cross-correlation coefficients. For the irreversible Michaelis-Menten scheme, for example, a similar analysis was carried out by Grima (2009). The present results should emerge as appropriate limiting cases from this more systematic procedure, which would also help us in establishing the precise regimes of their validity (e.g., the conditions on the turnover rates).

In summary, in this paper, we have analyzed ZOU mathematically under the tQSSA in the linear noise regime and derived explicit mathematical expressions for the averages and fluctuations. We showed rigorously that the critical point for the transition is unaffected by the inclusion of intermediate complexes (see also Xu and Gunawardena (2012)) and is robust against fluctuations. The variance of the modified substrate fraction shows a power-law divergence in the vicinity of the critical point, similar to the divergence of magnetic susceptibility near the second order paramagnet-ferromagnet phase transition. Beyond this similarity, however, the relation between ZOU and a phase transition remains obscure because (a) there is no direct interaction between substrate molecules, and therefore no diverging correlation length close to criticality (b) the transition from one phase to another is switch-like rather than continuous.

In recent times, there has been a renewed interest in various aspects of Michaelis-Menten kinetics, especially considering the widespread use of these rates in modern systems biology. ZOU occupies a special place among the various phenomena associated with enzyme-controlled post-translational modification of proteins, and has been the subject of a fair number of papers following the theoretical study of Goldbeter and Koshland (1981). However, as far as mathematical and computational studies are concerned, few papers venture beyond the standard mean-field framework, characterized by chemical rate equations. It is, nevertheless, a fact that protein copy numbers in biological cells are small on the scale of the Avogadro number, the magnitude of which provides the justification for neglect of fluctuations in most of standard thermodynamics. The LNA used here, under the simplifying conditions of tQSSA, was found to be successful in reproducing the fluctuations observed in simulations at high enough substrate concentrations, but significant differences were observed at lower values (corresponding to ~10,000 substrate proteins or less). Corrections to LNA are therefore, likely to be important in biological situations, and should be explored within the broader framework of multivariate system size expansion (van Kampen, 2008). We hope that our study will stimulate further work in this direction.

## Acknowledgements

This work was supported financially through a Fast-track project SR/FTP/PS-18/2010 funded by the Science and Engineering Research Council (SERC) of the Ministry of Science and Technology, Government of India. We also thank the P.G. Senapathy Centre for Computing Resources at IIT Madras for computing time in the Virgo cluster.

## Appendix A

Expansion of the Master Equation in inverse powers of the number of molecules

Defining $x = m_1/M_1$, $\delta_x = M_1^{-1}$ and $\Phi_x(x) \cong \delta_x^{-1} P_x(m_1)$, the master equation (1) becomes

$$\frac{\partial \Phi_x(x,t)}{\partial t} = \omega_+^x(x-\delta_x)\Phi_x(x-\delta_x) + \omega_-^x(x+\delta_x)\Phi_x(x+\delta_x) - [\omega_+^x(x) + \omega_-^x(x)]\Phi_x(x), \quad (A.1)$$

where, $\omega_+^x(x) = M_1(1-x)k_+ R_f$ and $\omega_-^x(x) = M_1 x k_-$. Now we expand the probability distribution function and the rates in a power series in $\delta_x$, i.e.,

$$\Phi_x(x \pm \delta_x) = \Phi_x(x) \pm \delta_x \Phi_x'(x) + \frac{\delta_x^2}{2} \Phi_x''(x) \pm O(\delta_x^3)$$

$$\omega_\pm(x \pm \delta_x) = \omega_\pm(x) \pm \delta_x \omega_\pm'(x) + \frac{\delta_x^2}{2} \omega_\pm''(x) \pm O(\delta_x^3) \quad (A.2)$$

The superscript $x$ on the rates is omitted for clarity in (A.2). Substitution of (A.2) into (A.1) yields, up to terms that are $O(\delta_x^2)$, the following partial differential equation:

$$\frac{\partial \Phi_x(x,t)}{\partial t} \approx \frac{\partial}{\partial x}\left[\delta_x\left(\omega_-^x(x) - \omega_+^x(x)\right)\Phi_x\right] + \frac{\partial}{\partial x}\left[\frac{\delta_x^2}{2}\frac{\partial}{\partial x}\left(\left(\omega_+^x(x) + \omega_-^x(x)\right)\Phi_x\right)\right]. \quad (A.3)$$

(A.3) has the same form as the equation of continuity (3) expressing the conservation of probability, with the probability current density given by (4a). Note that (A.3) is a Smoluchowski (Fokker-Planck) equation for $\xi$ with drift and diffusion coefficients given by (4b).

## Appendix B

Derivation of the GK cubic equation

In the sQSSA scheme, the concentration of the intermediates, denoted by $\tilde{A}$ and $\tilde{A}^*$ respectively are first expressed in terms of the concentrations of the unmodified ($A$) and modified substrate ($A^*$) as follows

$$\tilde{A} = \frac{k_+ R_0 A}{v_r + k_- + k_+ A} \quad (B.1)$$

$$\tilde{A}^* = \frac{k_+' B_0 A^*}{v_b + k_-' + k_+' A^*} \quad (B.2)$$

Next, using the steady state condition $v_r \tilde{A} = v_b \tilde{A}^*$ obeyed by the complete two state system, we obtain a relation between $A$ and $A^*$:

$$A^* = \frac{(v_b + k'_-)A}{\lambda(v_r + k_-) + (\lambda k_+ - k'_+)A} \quad \text{where} \quad \lambda = \frac{v_b k'_+ B_0}{v_r k'_- R_0} \tag{B.3}$$

Finally, by incorporating (B.1), (B.2) and (B.3) in the conservation relation $A + \tilde{A} + A^* + \tilde{A}^* = A_0$, we arrive at the following cubic equation in A:

$$k_+(\lambda k_+ - k'_+)A^3 + [k_+(\lambda(v_r + k_-) + (v_b + k'_-)) + (\lambda k_+ - k'_+)(v_r + k_- + k_+ R_0(1+v)) - A_0 k_+)]A^2$$
$$+ (v_r + k_-)[(\lambda(v_r + k_- + k_+ R_0(1+v)) + (v_b + k'_-) - A_0(2k_+\lambda - k'_+)]A - \lambda A_0(v_r + k_-)^2 = 0 \tag{B.4}$$

After obtaining the solution to (B.4) by numerical means, and using (B.2) and (B.3), the average modified fraction is given by

$$\bar{\xi} = (A^* + \tilde{A}^*)/A_0 \tag{B.5}$$

as a function of $R_0$. We may also obtain the expressions for fractions of unmodified ($x$) and modified ($y$) intermediate complexes, as functions of $R_0$:

$$x = \frac{\tilde{A}}{A + \tilde{A}} \quad \text{and} \quad y = \frac{\tilde{A}^*}{\tilde{A} + \tilde{A}^*} \tag{B.6}$$

**Appendix C**

Relation to the analytical results in Berg et al. (2000)

It is, naturally, interesting to explore the connections between our results and the analytical results in Berg et al. (2000), where explicit results for the probability distribution of the modified fraction, as well as the first and second moments were computed. Following our arguments in the section `Models and Methods', we find that the steady state distribution $P(\xi)$ (see (6)) is given by

$$P(\xi) \propto \frac{1}{\bar{D}_\xi(\xi)} \exp\left(\int_0^\xi \frac{\bar{v}_\xi(\xi')}{\bar{D}_\xi(\xi')} d\xi'\right), \tag{C.1}$$

where the coefficients $\bar{v}_\xi$ and $\bar{D}_\xi$, in the rational approximation, are given by (16) and (30). In the limit where $A_0$ is large, $r, \kappa_r, b, \kappa_b \to 0$, leading to

$$\bar{v}_\xi \sim v_r V(R_0 - R_c) \quad \text{and} \quad \bar{D}_\xi \sim \frac{v_r}{2A_0}(R_0 + R_c). \tag{C.2}$$

Therefore, from (41),

$$P(\xi) \propto \exp\left[2N\left(\frac{R_0 - R_c}{R_0 + R_c}\right)\xi\right] \quad \text{as} \quad A_0 \to \infty.$$

(C.3)

Note that the corresponding expression in Berg et al. (2000), in our notation, is $P(n) \mu (R_0/R_c)^n$ with $n = N\xi$, which is similar to (B.3) after exponentiation; the apparent difference may be attributed to the discrete treatment in Berg et al. (2000), compared to the continuum approach followed in the present paper. It can be shown rigorously that the expressions become identical as $R_0 \to R_c$.

The mean and variance of $\xi$, in the present approximation, are easily computed from (C.3), and the results are

$$\bar{\xi} = 1 - \frac{(R_0 + R_c)}{N(R_0 - R_c)} \quad \text{when} \quad R_0 > R_c \quad \text{and} \tag{C.4a}$$

$$\bar{\xi} = \frac{(R_0 + R_c)}{N(R_c - R_0)} \quad \text{when} \quad R_0 < R_c \quad \text{and} \tag{C.4b}$$

$$\bar{\xi} = \frac{1}{2} \quad \text{when} \quad R_0 = R_c. \tag{C.4c}$$

Similarly, the variance is given by

$$\sigma_\xi^2 = \frac{1}{N}\left[\frac{R_0 + R_c}{R_0 - R_c}\right]^2 \quad \text{when} \quad R_0 \neq R_c \quad \text{and} \tag{C.5a}$$

$$\sigma_\xi^2 = \frac{1}{12} \quad \text{when} \quad R_0 = R_c. \tag{C.5b}$$

Not surprisingly, the expressions in (C.4c) and (C.5b) match with the corresponding results in Berg et al. (2000). However, both disagree with (27) and (37c) respectively, which points to the inherent weakness of this approach. See also simulation data in Fig. **6**

| Parameter | Numerical value |
|---|---|
| $A_0$ | 5.3-54.4 $\mu M$ |
| $B_0$ | 2.28 $\mu M$ |
| $k_+$ | 25.64 $\mu M^{-1} s^{-1}$ |
| $k_-$ | $10 s^{-1}$ |
| $k'_+$ | 18.51 $\mu M^{-1} s^{-1}$ |
| $k'_-$ | $10 s^{-1}$ |
| $v_r$ | 0.75 $s^{-1}$ |
| $v_b$ | 0.6 $s^{-1}$ |
| $V$ | $10^{-18} m^3$ |

**Table 1.** A list of the values of parameters used in numerical simulations. For specificity and in accordance with our goals for the future, we have generally chosen values relevant to the chemotaxis network of *E. coli* (see Emonet and Cluzel (2008), and references therein). The off-rates $k_-$ and $k'_-$ have been arbitrarily assigned a value of $10 s^{-1}$ (much larger than $v_r$ and $v_b$); $k_+ = k_-/K_r$ and $k'_+ = k'_-/K_b$ are then fixed using the experimentally measured dissociation constants $K_r = 0.54 \mu M$ and $K_b = 0.39 \mu M$. The lowest value of $A_0$ is closer to [CheA] found in *E. coli*, and higher values were simulated to see the emergence of ultrasensitivity. The value of $B_0$ is higher than what was used in Emonet and Cluzel (2008), see Reneaux and Gopalakrishnan (2010) for a brief discussion of this point. The GK point for these parameters is $R_c = 1.824 \mu M$.

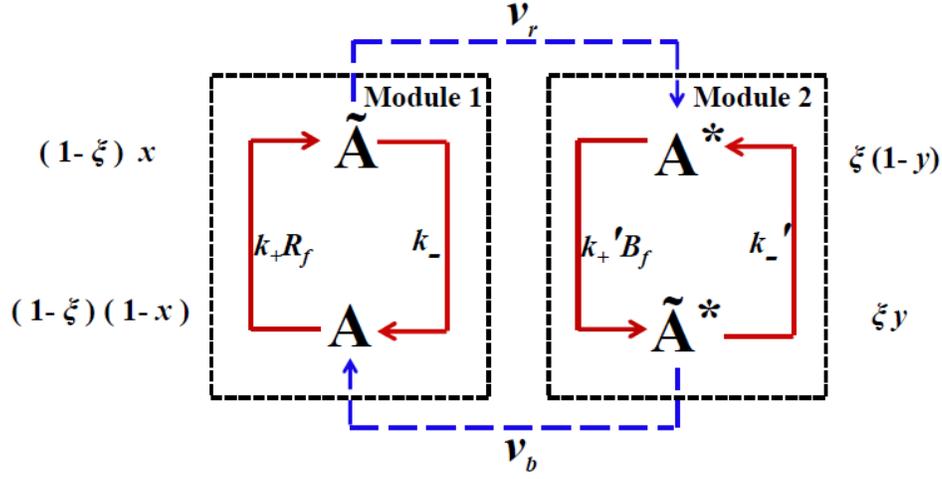

**Fig. 1.** A schematic diagram depicting the various dynamical processes, as well as the modules in our system. The expressions on the left and right sides denote the fractional concentrations of the corresponding species.

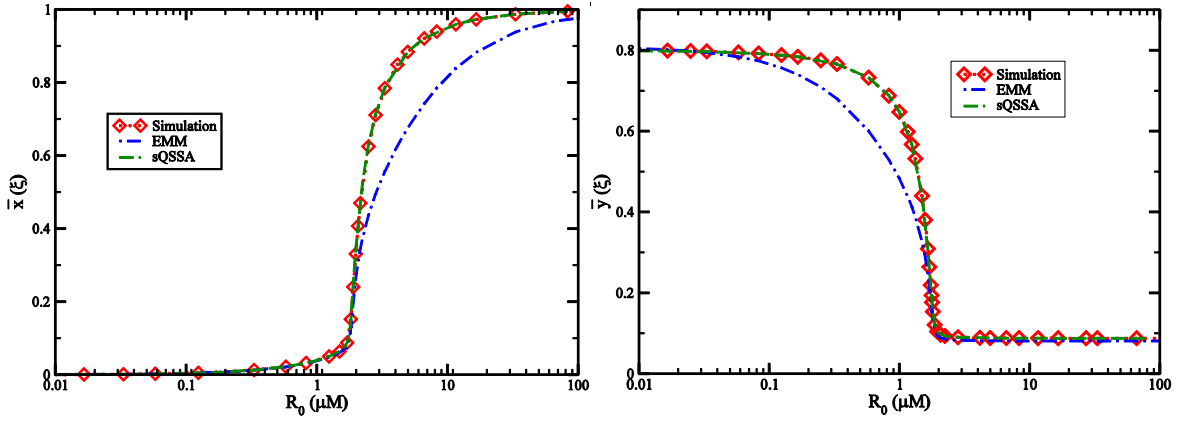

**Fig. 2.** A comparison between the numerical results for $\bar{x}(\xi)$ and $\bar{y}(\xi)$, the corresponding EMM expressions from (31a) and (31b) and the GK results (B.6) for $A_0 = 27.2\,\mu M$

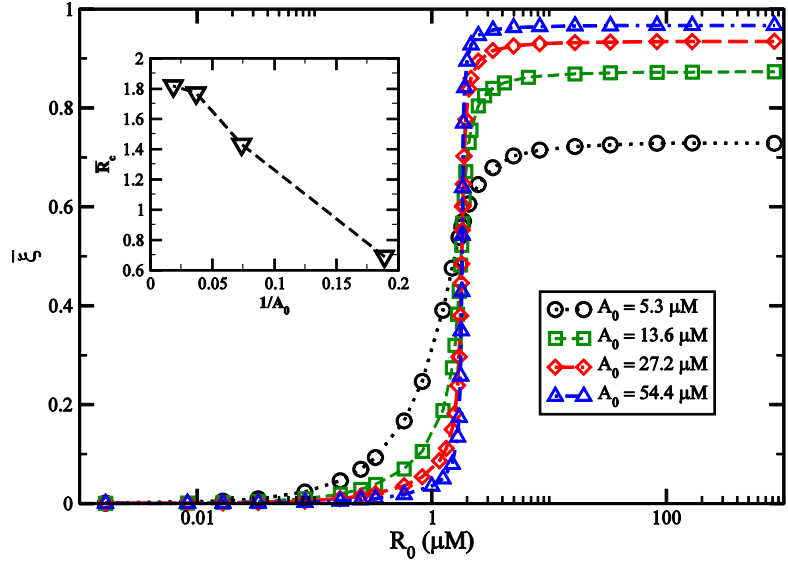

**Fig. 3a.** The figure shows $\bar{\xi}$ as a function of $R_0$ at fixed $B_0$ for various $A_0$. Inset: The effective critical point $\bar{R}_c(A_0) \equiv \int_0^\infty dR_0 \left(\partial \bar{\xi} / \partial R_0 \right)$ is plotted against $A_0^{-1}$, showing convergence to the GK point as $A_0^{-1} \to 0$.

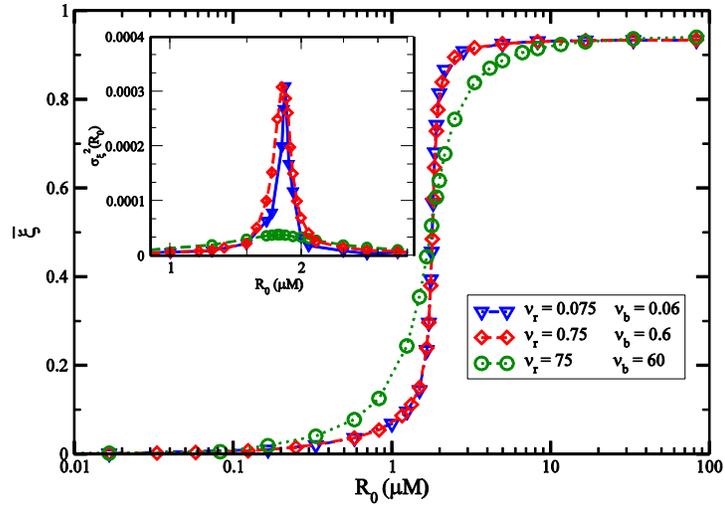

**Fig 3b.** The figure shows the simulation results ($A_0 = 27.2 \mu M$) for the ZOU transition for different values of the final turnover rates, keeping, however, the same ratio between them. ZOU is weakened when the rates $v_r$ and $v_b$ are increased. Inset: Reduction in fluctuations also confirms this effect

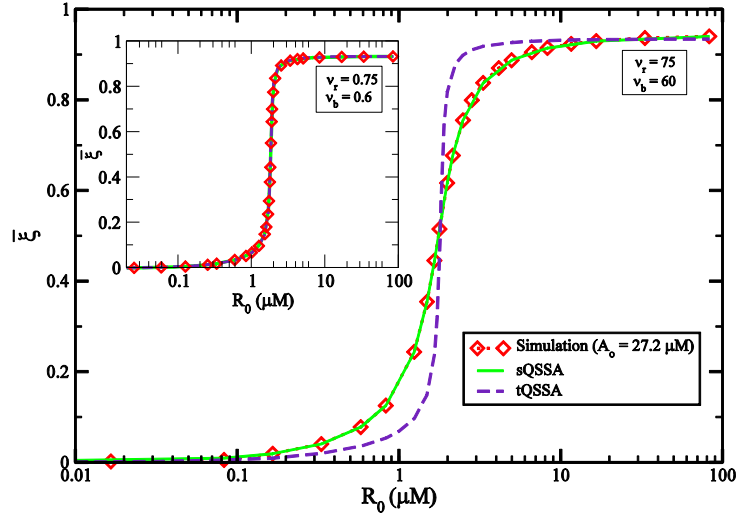

**Fig. 4a** Comparison of theoretical curves predicted by (a) the van Kampen fixed point equation (17), denoted tQSSA and (b) the Goldbeter-Koshland cubic equation, denoted sQSSA ((B.4) and (B.5)), for two different regimes of transition rates. As expected, at high $v_r$ and $v_b$ values, the latter gives a better estimate, while for lower values (inset), both of them agree perfectly with simulation.

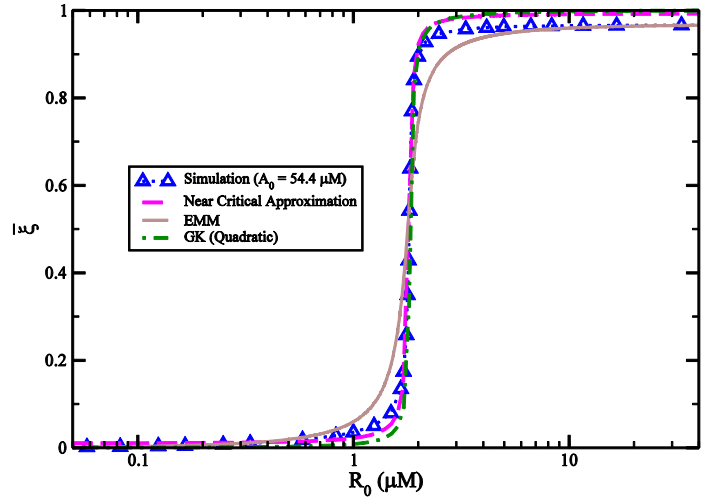

**Fig. 4b.** The mean total substrate fraction $\bar{\xi}$ as a function of $R_0$ at $A_0 = 54.4 \mu M$ as obtained from (a) numerical simulations, (b) near-critical approximation under tQSSA (25b) (b) the Goldbeter-Koshland curve under sQSSA with intermediates neglected (28b) and (c) the EMM approximation (33b).

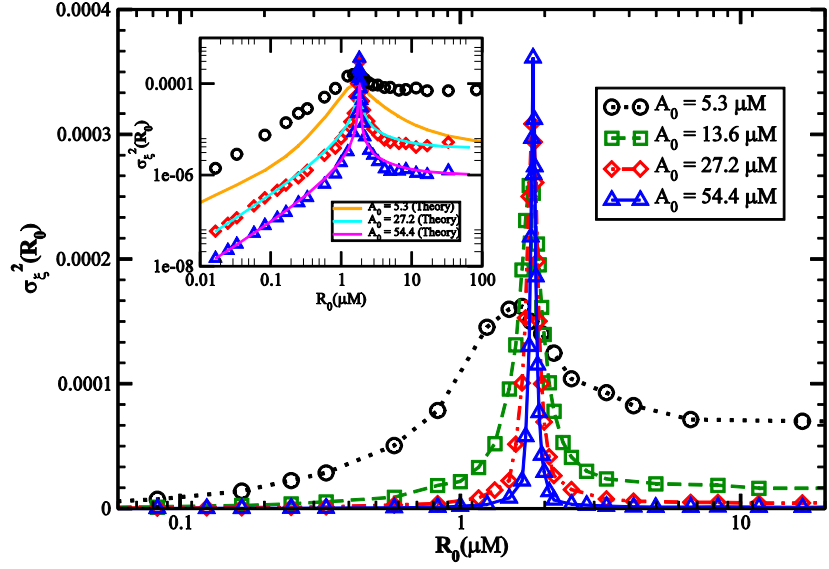

**Fig. 5.** The figure clearly shows that the variance in $\xi$ is maximized at the critical point. Inset: Selected data is shown on a logarithmic scale, along with a <u>fitting curve</u> derived under the EMM approximation (38), combined with $\bar{\xi}$ as measured in simulations (data shown in Fig. 3a). Note that the data for the smallest substrate concentration deviates significantly from the theoretical prediction, underscoring the limitations of LNA.

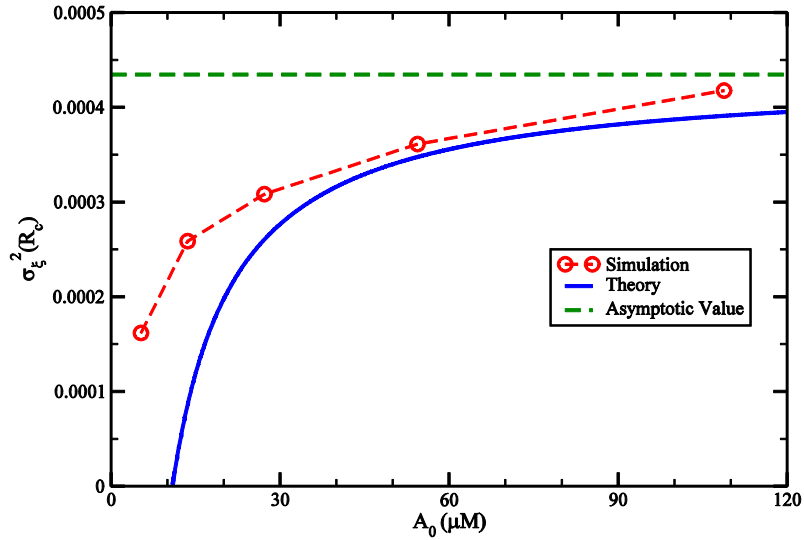

**Fig. 6.** The critical variance obtained from simulations is plotted against $A_0$, along with a comparison with the analytical expression obtained by substituting (26a) and (26b) in (36). The asymptotic value predicted by (37c) is also indicated in the figur

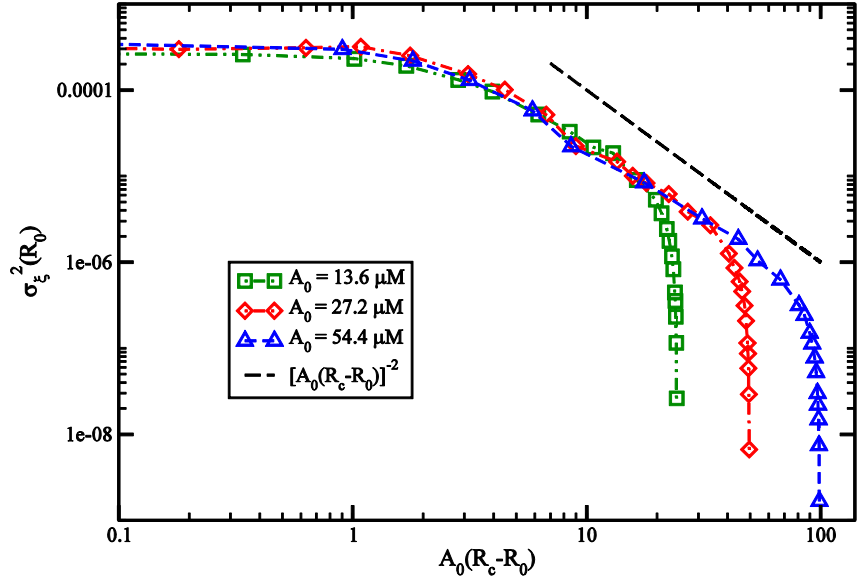

**Fig. 7a.** The simulation results for variance at different $A_0$ values show good data collapse with a clear evidence for the scaling behavior of the variance in the sub-critical regime, as predicted by (37a), with the scaling function displaying an algebraic decay with exponent -2.

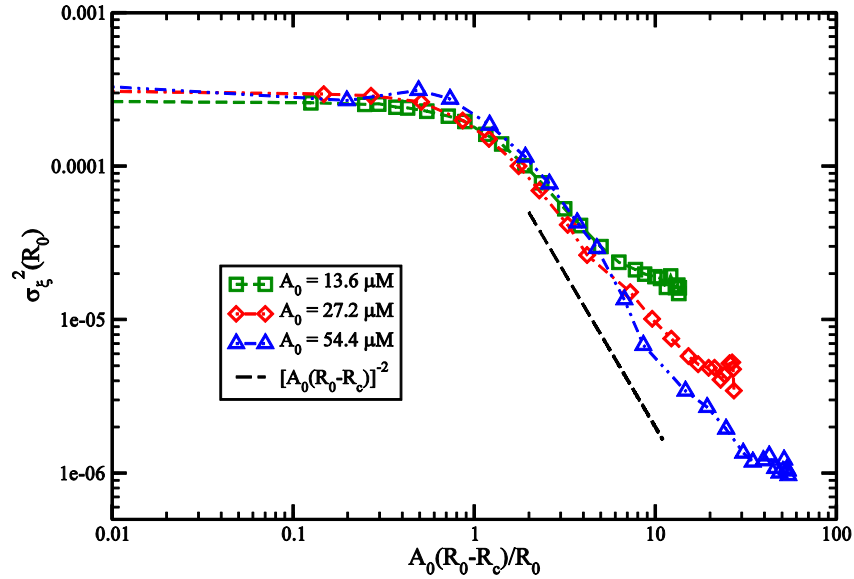

**Fig. 7b.** The scaling behavior of the variance, in the super-critical regime. The scaling factor in the x-axis is suggested by (37b), where we have approximated $R_0 + R_c \approx 2R_c$ in the numerator. Admittedly, the scaling collapse here is not as good as what is found in the sub-critical regime (Fig. 7a).

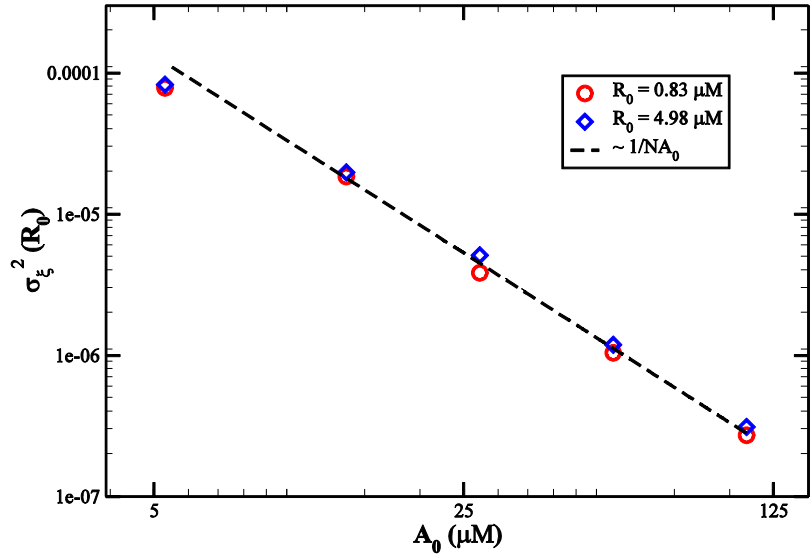

**Fig. 8.** The figure shows that, away from the critical point, the fluctuations in $\xi$ reduce with increase in $A_0$, unlike its behavior at the critical point itself. The sub-critical and super-critical points falling on top of each other is accidental.